\documentclass{aa}  

\usepackage{graphicx}
\usepackage{txfonts}
\usepackage{upgreek}
\usepackage{gensymb}
\usepackage{natbib}
\newcommand{\mg}{MG~0751+2716}
\newcommand{\class}{CLASS~B1600+434}
\usepackage{bm}

\usepackage[dvipsnames]{xcolor}

\begin{document}

\title{High-resolution imaging with the International LOFAR Telescope: Observations of the gravitational lenses \mg\ and CLASS~B1600+434}

\author{Shruti Badole\inst{1}, Deepika Venkattu\inst{1,2}, Neal Jackson\inst{1}, Sarah Wallace\inst{1}, Jiten Dhandha\inst{1}, Philippa Hartley\inst{1,3}, Christopher Riddell-Rovira\inst{1}, Alice Townsend\inst{1}, Leah K. Morabito\inst{4,5}, J. P. McKean\inst{6,7}
}

\institute{
Jodrell Bank Centre for Astrophysics, Department of Physics and Astronomy, University of Manchester, Oxford Rd, Manchester M13 9PL, UK\\ \email{shruti.badole@postgrad.manchester.ac.uk}
\and Department of Astronomy and The Oskar Klein Centre, AlbaNova University Center, Stockholm University, SE-106 91 Stockholm, Sweden
\and Square Kilometre Array Organisation, Jodrell Bank, Lower Withington SK11 9FT, Cheshire, UK
\and Centre for Extragalactic Astronomy, Department of Physics, Durham University, Department of Physics, South Road, Durham DH1 3LE, UK
\and Institute for Computational Cosmology, Department of Physics, Durham University, South Road, Durham DH1 3LE, UK
\and ASTRON, Netherlands Institute for Radio Astronomy, Oude Hoogeveensedijk 4, Dwingeloo, the Netherlands
\and Kapteyn Astronomical Institute, University of Groningen, P.O. Box 800, 9700AV Groningen, the Netherlands
\\
}
\titlerunning{LOFAR observations of gravitational lenses}
\authorrunning{Badole et al.}
\date{Received (date); accepted (date)}

 
\abstract{We present  Low-Frequency Array (LOFAR) telescope observations of the radio-loud gravitational lens systems \mg\ and \class . These observations produce images at 300 milliarcseconds (mas) resolution at 150~MHz. In the case of \mg, lens modelling is used to derive a size estimate of around 2~kpc for the low-frequency source, which is consistent with a previous 27.4 GHz study in the radio continuum with Karl G. Jansky Very Large Array (VLA). This consistency implies that the low-frequency radio source is cospatial with the core-jet structure that forms the radio structure at higher frequencies, and no significant lobe emission or further components associated with star formation are detected within the magnified region of the lens.  \class\ is a two-image lens where one of the images passes through the edge-on spiral lensing galaxy, and the low radio frequency allows us to derive limits on propagation effects, namely scattering, in the lensing galaxy. The observed flux density ratio of the two lensed images is 1.19$\pm$0.04 at an observed frequency of 150~MHz. The widths of the two images give an upper limit of 0.035~kpc\,m$^{-20/3}$ on the integrated scattering column through the galaxy at a distance approximately 1~kpc above its plane, under the assumption that image A is not affected by scattering. This is relatively small compared to limits derived through very long baseline interferometry (VLBI) studies of differential scattering in lens systems. These observations demonstrate that LOFAR is an excellent instrument for studying gravitational lenses. We also report on the inability to calibrate three further lens observations: two from early observations that have less well determined station calibration, and a third observation impacted by phase transfer problems.}
  

\maketitle

\section{Introduction}

Strong gravitational lenses are systems in which a background galaxy is multiply imaged by the gravitational field of a foreground galaxy- or cluster-scale mass, typically at redshift $z\sim 0.5-1$, and typically producing two or four images of the background source (see \citealt{2010ARA&A..48...87T} and \citealt{2010CQGra..27w3001B} for reviews). They are astrophysically useful tools, for three main reasons. Firstly, they can be used to determine a very accurate total mass within the Einstein radius of the lens galaxy, and they can constrain its mass distribution (e.g.  \citealt{1995ApJ...447...62C,1997MNRAS.292..148S,2001ApJ...554.1216C,2002MNRAS.330..205R,2003ApJ...583..606K,2009MNRAS.398..607V,2012ApJ...750...10S}).  Secondly, lensing preserves surface brightness, so the effective increase in solid angle containing images of the source allows us to investigate the background source at some combination of higher resolution and higher sensitivity (e.g. \citealt{1989MNRAS.238...43K,2003ApJ...590..673W,2018MNRAS.478.4816S,2018MNRAS.476.4383D,2019MNRAS.485.3009H,2020MNRAS.494.5542R,2020MNRAS.496..138B}). Finally, gravitational lens systems may be used to investigate environmental effects in the lensing galaxy, since lens systems consist of multiple images of the same object seen along different lines of sight through the lensing galaxy. This can be done in a number of ways, including differential effects on radio polarization, differential optical extinction, differential scattering of radio waves, or differential X-ray absorption (e.g. \citealt{2003A&A...405..445W,2003MNRAS.338..599B,2006ApJS..166..443E,2009ApJ...692..677D,2011ApJ...728..145M,2017NatAs...1..621M}).

Several hundred gravitational lens systems are now known, of which a minority (approximately 10\%) contain radio-loud background sources, either in the form of radio-loud quasars, which produce compact images of the radio core and jet emission (e.g. \citealt{2004MNRAS.350..949B}), or in extended images of large-scale radio jets. In addition, many of the 90\% of radio-quiet lens systems have detectable radio emission at the level of a few tens of \textmu Jy, although in many cases this is likely to be produced by star-forming processes \citep{2008evn..confE.102W,2011ApJ...739L..28J,2015MNRAS.454..287J,2020MNRAS.496..138B}.

The first aim of this work is to use lens models to reconstruct the radio source. Many studies of radio lenses have attempted to do the same (e.g. \citealt{1989MNRAS.238...43K,1993AJ....105..847L,1997MNRAS.289..450K,2004MNRAS.349....1W,2004MNRAS.349...14W,2019MNRAS.485.3009H,2020MNRAS.495.2387S,2020MNRAS.493.5290S}) and study the relation of the radio source components to other components of the source emitting at other wavebands. The relation of steep-spectrum, synchrotron radio components with other source components can also be investigated, given sufficient resolution at low frequency, which no instrument other than the Low-Frequency Array (LOFAR) operated by the International LOFAR Telescope (ILT) foundation is capable of providing. 

A second aim of this work is to use radio lenses to explore environmental effects in the lens galaxy. Optical studies have shown effects including reddening of lensed images passing through the lens galaxy disk \citep{2006ApJS..166..443E,2011ApJ...742...67M}, and microlensing, which can reveal the lens galaxy stellar population (for a review, see e.g. \citealt{2012RAA....12..947M}). A number of the known radio lenses show significant foreground effects, notably CLASS~B0218+357, which has a range of absorption effects due to passage through molecular gas in the spiral lensing galaxy \citep{1995A&A...299..382W,1996ApJ...465L..99M, 2006A&A...447..515M, 2007A&A...465..405M}, and PKS1830$-$211, in which absorption effects are also seen \citep{1996Natur.379..139W,1998ApJ...500..129W}. In addition to  absorption, radio waves may be scattered by ionised columns within any intervening object. Very long baseline interferometry (VLBI) studies that show broadening of one or more components are indicative of scattering either in the lensing galaxy \citep{1999MNRAS.305...15M,1996ApJ...470L..23J,2003MNRAS.338..599B,2003ApJ...590...26W,2004MNRAS.350..949B} and occasionally in our own Galaxy \citep{2003ApJ...595..712K}. Scattering effects are more noticeable at lower frequencies, generally increasing as $\nu^{-2}$, and therefore should be more prevalent, provided high enough resolution (which only LOFAR can provide) can be obtained to separate the lensed images and quantify any scatter-induced broadening in them. Such high levels of scattering or free-free absorption could also impact potential wide-area surveys for lenses at low radio frequencies.

A few  high-resolution observations of gravitational lenses have previously been made at frequencies below 500~MHz. These have included low-frequency VLBI observations at 327~MHz (e.g. \citealt{2008ApJ...673...78L}) and Multi-Element Radio Linked Interferometer Network (MERLIN) observations at 408~MHz (e.g. \citealt{1980Natur.288...69N}). The advent of LOFAR \citep{2013A&A...556A...2V} at still lower frequencies, however, has opened up a new spectral window for high-sensitivity observations at low radio frequencies. Recent progress in the  calibration of the LOFAR international baselines \citep{paper1} offers the prospect of routine studies at 30-200~MHz with resolutions of 200-300 milliarcseconds (mas), ideally matched to the angular scale of the images in strong gravitational lens systems, which typically have image separations of the order of 1$^{\prime\prime}$. LOFAR observations, given their low radio frequency, are uniquely able to probe steep-spectrum synchrotron-emitting plasma at resolutions comparable to Atacama Large Millimeter Array (ALMA) or the Very Large Array (VLA) at THz and GHz frequencies, respectively. At these low radio frequencies, we also gain a significant advantage in the study of scattering and related phenomena, given the $\nu^{-2}$ dependence of scattering and free-free absorption effects.

As a first step, we present LOFAR maps produced using observations that included the international baselines of two lens systems. We first investigate \mg\ \citep{1988Natur.333..537H} in order to study the lensed source, to deduce its overall size and to compare it to that observed at shorter wavelengths. We also present an observation of \class\ \citep{1995MNRAS.275.25-29}; this lens system consists of a quasar that is lensed by an edge-on spiral galaxy. Because of the passage of one of the images close to the plane of the galaxy, this offers an opportunity to investigate a system in which scattering effects and absorption would be potentially important. Lastly, we also discuss three observations that we were unable to calibrate successfully.

Where necessary, we assume a standard flat Universe with $H_0=70$\,km\,s$^{-1}$\,Mpc$^{-1}$ and $\Omega_{\Lambda}$= 0.7.

\section{\mg}

\subsection{The lens system \mg}

\mg\ (07$^{\rm h}$51$^{\rm m}$41.5$^{\rm s}$ 27$^{\circ}$16$^{\prime}$31$^{\prime\prime}$) is a gravitational lens system originally observed in the MIT-Greenbank radio survey \citep{1990ApJS...72..621L} and identified as a gravitational lens by \cite{1997AJ....114...48L}. It consists of a $z=3.2$ quasar lensed by a galaxy at redshift 0.35 \citep{1999AJ....117.2034T}. It is a bright radio source, with a flux density of  1.47~Jy at 365~MHz \citep{1996AJ....111.1945D}. Its radio spectrum peaks at a few hundred MHz, but is steeper at higher frequencies, and it consists of lensed jet emission from a classical synchrotron radio source originating in an active galactic nucleus (AGN). The core-jet system is gravitationally lensed, resulting in a complex structure stretched out along an arc of emission \citep{1997AJ....114...48L}. In addition to being a source of  strong radio emission, the quasar was discovered to  be a source of CO molecular line emission  \citep{2002A&A...385..399B,2007A&A...470...53A,2011ApJ...739L..32R,2020MNRAS.495.2387S} as well as sub-millimetre continuum \citep{2009ApJ...707..988W,2018MNRAS.476.5075S}. 

Its radio structure was mapped in detail by \cite{2018MNRAS.478.4816S}, who studied \mg\ using 1.65 GHz global VLBI observations and obtained images at milliarcsecond resolution. They found  evidence of  low-mass structure in the mass distribution of the lens, although it is not certain whether this is in the form of $10^6$ to $10^8$~M$_{\odot}$ sub-haloes or more complex mass distributions in the group associated with the main lensing galaxy. Here we use the observed structure at a frequency $\sim$10 times lower, together with lens modelling, to investigate whether
the  size of the lensed structure   at the lower frequency is similar to that at higher (GHz) frequencies.

\subsection{Observations and data reduction}

\mg\ was observed on 2018 January 17 with LOFAR as part of programme LC9-012 (PI: Jackson), using the High Band Array (HBA), with the HBA Dual Inner antenna set \citep{2013A&A...556A...2V} with a bandwidth ranging from 120 to 183~MHz. The observation was preceded by a ten-minute observation of the bright calibrator source 3C~196, and followed by a similar observation of 3C~295. This is the standard observing strategy for the LOFAR Two-metre Sky Survey (LoTSS; \citealt{2017A&A...598A.104S}). All stations were used, including all 13 international stations (6 in Germany, 3 in Poland and 1 each in Ireland, France, Sweden, and the UK). Data were initially recorded at 64 channels per 196 kHz sub-band, which was averaged to 16 channels per sub-band, and were recorded with one-second integration time.

The data analysis was performed during the development of the LOFAR-VLBI Pipeline described by \cite{paper1}. The initial calibration of the Dutch part of the array, including the core stations close to the centre of the array at Exloo, Netherlands, and the remote stations at baselines up to 80~km, was done using a model for 3C295 that was established for the LOFAR HBA by F. Sweijen\footnote{ \url{https://github.com/lofar-astron/prefactor}}. Good solutions for total electron content (TEC) and bandpass were obtained for all core and remote stations, using CS001 as the reference station, with the standard procedures (known as {\sc prefactor}\footnote{\url{https://www.astron.nl/citt/prefactor/}}, \citealt{2019A&A...622A...5D}). An initial phase calibration of the Dutch stations was then performed using an input sky model from the Tata Institute of Fundamental Research - Giant Metrewave Radio Telescope Sky Survey (TGSS) Alternative Data Release \citep{2017A&A...598A..78I}. Data at  frequencies higher than 166~MHz were excluded from the sample, due to severe radio frequency interference. 

\begin{figure}
    \includegraphics[width=8cm]{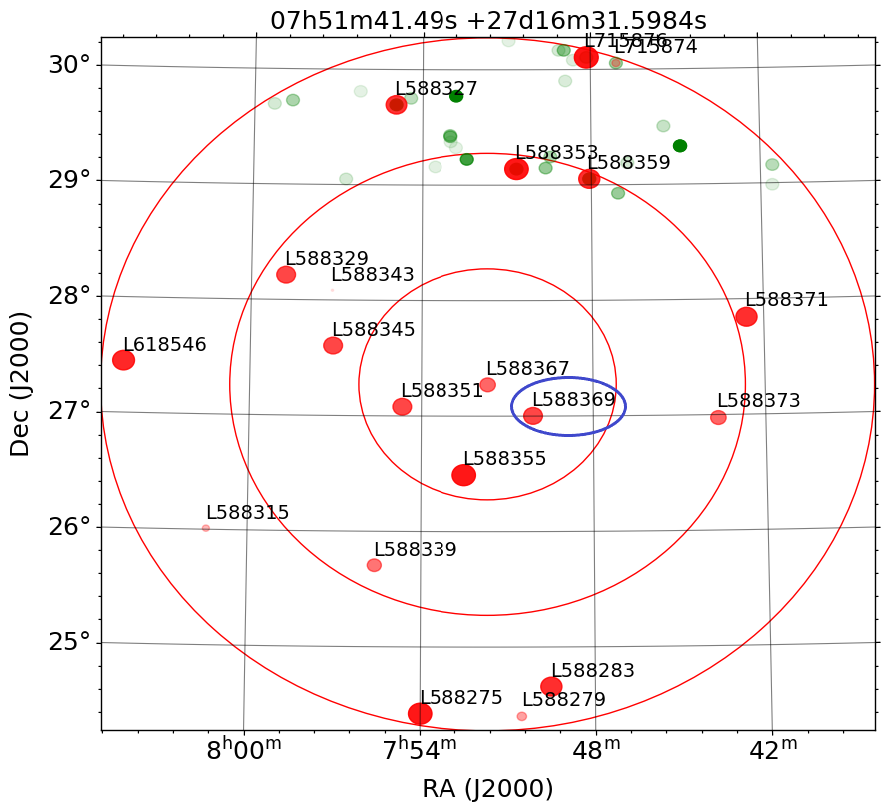}
    \caption{LBCS sources in the vicinity of MG 0751+2716. Sources from the Westerbork Northern Sky Survey (declination limit of about 29$^{\circ}$) are in green and LBCS sources are in red. The source indicated by the blue ellipse   is the phase calibrator L588369.}
    \label{lbcs0751}
\end{figure}

The LOFAR Long-Baseline Calibrator Survey (LBCS, \citealt{2016A&A...595A..86J,paperlbcs}) was used to search for phase calibrators near  the source (Fig.~\ref{lbcs0751}). A bright compact calibrator, L588369 (B0747+27), lies about half a degree away from MG 0751+2716, which itself is an LBCS source (L588367). This source was used to correct the clock (non-dispersive) and ionospheric (dispersive) delays, as outlined by \cite{paper1}. We solved for TEC using the LOFAR-VLBI Pipeline. Smoothly varying solutions with most values lying below 1~TECU (1 TECU = $10^{16}$ electrons/$\mathrm{m^2}$) were found for all international stations, with gradients of at most 1~TECU/hour.

Once the delay solutions were   obtained, they were applied to the data. The entire dataset was then phase-rotated to the positions of the target and nearby calibrator in order to produce two smaller datasets, averaged by a factor of 8 in both time and frequency to give 8s integrations and channels of 97~kHz width. The Dutch core stations, which are all within approximately 4 kilometres of the centre of the array, were combined to form a single super-station.

Imaging and self-calibration was performed in Difmap \citep{1997ASPC..125...77S}, with both the phase calibrator source and target data initially phase self-calibrated using a point-source starting model. Although the phase calibrator source was intended for use in generating phase solutions to apply to the target, this proved unnecessary as the convergence of the target model was good; however, the phase calibrator was imaged using the same procedure as \mg\ to verify that the pipeline was working correctly. Six iterations of CLEANing were performed, using uniform weighting, 50 mas pixels, and a u-v plane taper of 50\% at 25k$\lambda$, in order to downweight the short baselines. Phase-only self-calibration was done between each CLEAN iteration, and final maps were produced with contour levels of three times the nominal r.m.s. noise level in the image close to the sources.

\begin{figure*}
    \begin{tabular}{cc}
    \includegraphics[width=8.8cm]{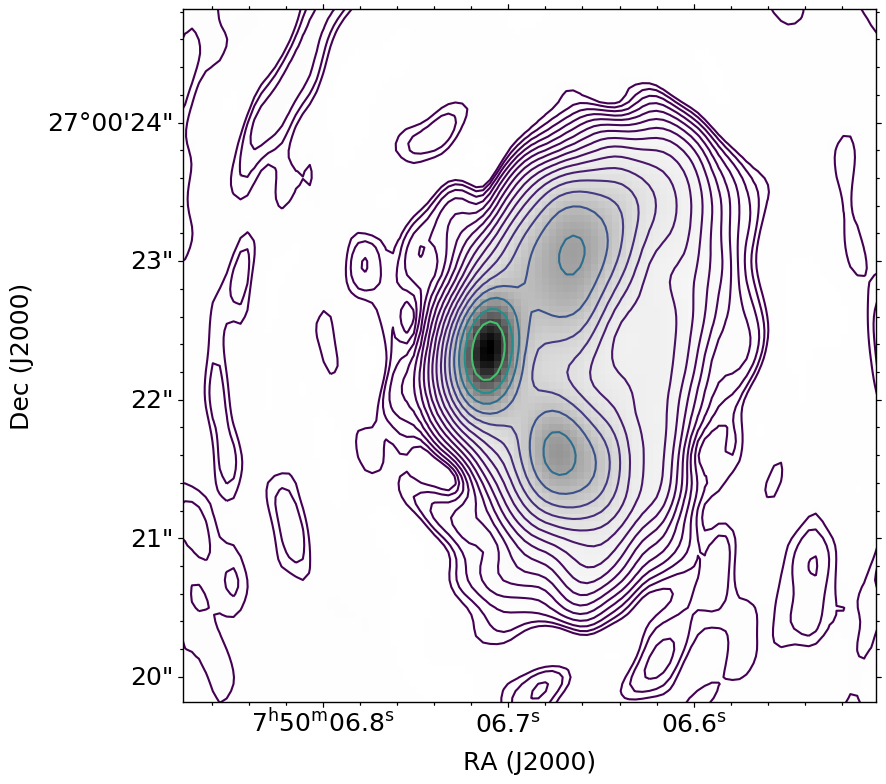}&
    \includegraphics[width=9.1cm]{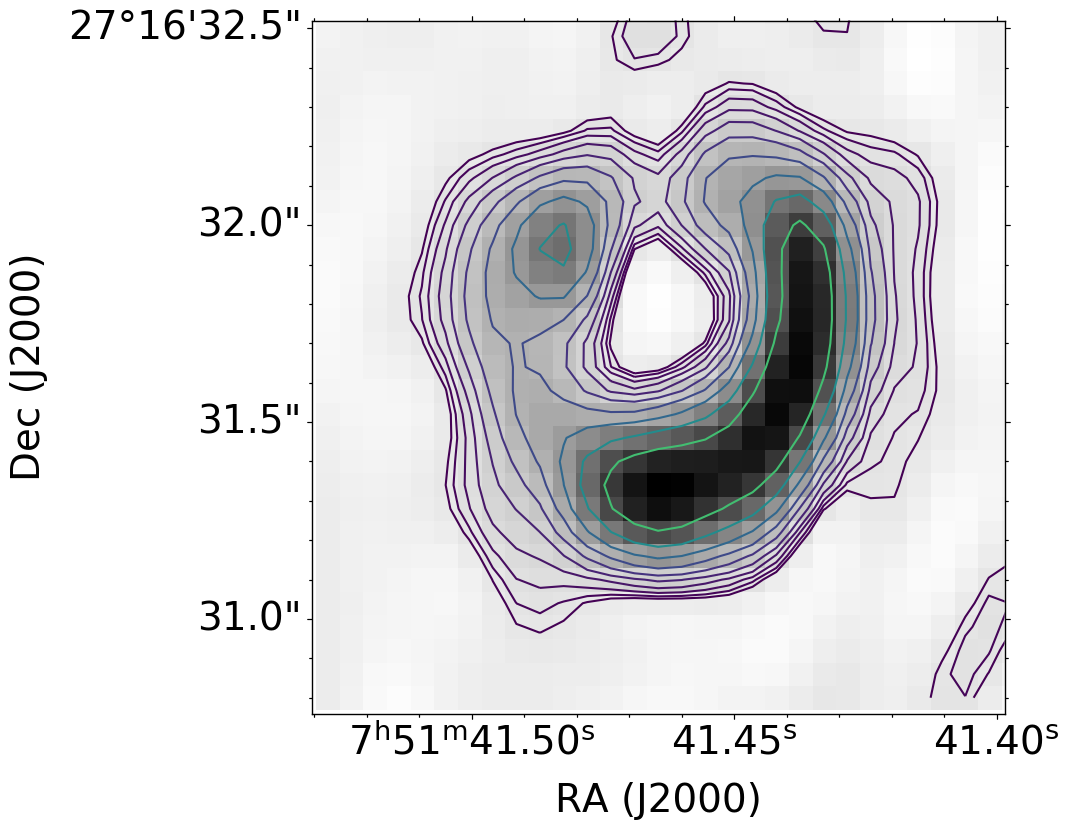}\\
    \end{tabular}
    \caption{LOFAR-HBA maps of L588369 and \mg. Left: Image of the phase calibrator L588369. The lowest contour is at 690~\textmu Jy/beam and contours increase by factors of $\sqrt{2}$. Right: Image of \mg. The lowest contour is at 790~\textmu Jy/beam and contours increase by factors of $\sqrt{2}$. The beam size is 0\farcs306 $\times$ 0\farcs184 with a position angle of $0.41^{\circ}$.}
    \label{lofar0751}
\end{figure*}

The maps of the calibrator source L588369 and the target source are shown in Fig.~\ref{lofar0751}. The calibrator source has a distorted structure of core and two lobes, possibly a small wide-angle tail source. MG~0751+2716 has the main structural features previously observed in investigations of this object, with a bright south-western arc and north-eastern counterimage
\citep{1997AJ....114...48L,2018MNRAS.478.4816S}.

The flux density scale in the maps of Fig.~\ref{lofar0751} was produced directly by {\sc prefactor}. The total flux density of the calibrator L588369 is measured as 1.4~Jy at a mean frequency of 150 MHz, which compares well to the 3.23~Jy at 74~MHz and 0.81~Jy at 408~MHz measured in the VLA Low-Frequency Sky Survey \citep{2007AJ....134.1245C} and B2 radio survey \citep{1972A&AS....7....1C}, respectively. Because the calibrator L588369 is about 0.5 degrees from the field centre, its flux density may be reduced by about 20-30\% by a combination of time and bandwidth smearing.

\subsection{Unlensed source modelling}

We incorporated the lens model, Model 1, described by \cite{2018MNRAS.478.4816S} (Table \ref{model1tab}) to fit the u-v data using the {\sc visilens} package \citep{hezaveh13a,spilker16a}. {\sc visilens} uses interferometric visibility data to arrive at the structure of the background source and the lens that form the gravitational lens system. The lens structure is described by a few key parameters, namely the critical radius of the lens, the position of the lens,
the ellipticity of the lens mass distribution, the position angle of the major axis, and the magnitude and position angle of the external shear. The critical radius of a lens is the radius of the Einstein ring for a spherical lens which is directly in front of the source and is proportional to the square root of the lensing galaxy mass. External shear refers to the distortion of the lensed images, in a preferred direction, caused by the lens galaxy environment.

\begin{table}[]
\centering
\caption{Parameters of the lens model, Model 1, determined by \citealt{2018MNRAS.478.4816S}, together with the range of the fitted values from the data in a procedure with Model 1 as a starting point.}
\begin{tabular}{lll}
\hline
\multicolumn{1}{l}{\textbf{Parameters}} & \multicolumn{1}{l}{\textbf{Lens model}} &
\multicolumn{1}{l}{\textbf{Fitted values}}\\ 
 \hline
{\em Lens parameters:} & ... & ... \\
$b ~(\arcsec)$    & 0.40249 &  0.403$\pm$0.007 \\
$\Delta x_L ~(\arcsec)$   & 0.052 & ... \\
$\Delta y_L ~(\arcsec)$  & 0.3804 & ...  \\
$e$             & 0.159   & ... \\
$\theta ~(^\circ)$     & 35.7     & ...       \\
$\Gamma$     & 0.0837   & ...   \\
$\Gamma_{\theta} ~(^\circ)$  & 79.2 & ... \\ & & \\
{\em Source parameters:} & ... & ... \\
$\Delta x_S ~(\arcsec)$ & ... & 0.041$\pm$0.022 \\
$\Delta y_S ~(\arcsec)$ & ... & $-$0.012$\pm$0.001 \\
Flux density (mJy) & ... & 7.1$\pm$1.3 \\
FWHM ~$(\arcsec)$& ... & 0.048$\pm$0.018 \\
Axis ratio & ... & 0.46$\pm$0.23 \\
Position angle ~$(^\circ)$& ... & $73.3\pm4.5$ \\ \hline
\end{tabular}
\tablefoot{$b$ is the critical radius of the lens galaxy, $\Delta x_L$ and $\Delta y_L$ are the positions of the lens (in right ascension and declination) with respect to the phase centre of the observations,  $\Delta x_S$ and $\Delta y_S$ are the fitted positions of the source with respect to the lens, $e$ is the ellipticity of the lens,  $\theta$ is the position angle of the ellipticity, $\Gamma$ is the external shear magnitude, and $\Gamma_{\theta}$ is the external shear position angle. All angles are in degrees, east of north.}
\label{model1tab}
\end{table}

The density slope $\gamma$ of the ellipsoidal power-law mass distribution in the original Model 1 is 2.079; however, we assumed a singular isothermal ellipsoid (SIE) profile for our lens model fitting, for which $\gamma=2$, and let the mass of the lensing galaxy vary. We also let all the source parameters vary, assuming the source to have a Gaussian profile. {\sc visilens} uses the package `emcee' to carry out a Markov chain Monte Carlo (MCMC) analysis \citep{2013PASP..125..306F}.

\begin{figure*}
    \includegraphics[width=19cm]{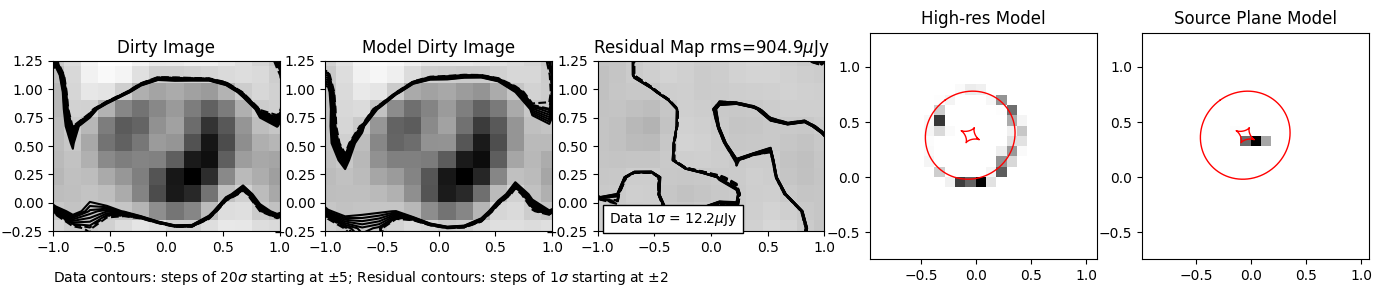}
    \caption{One of the results of the various MCMC  runs to fit a source to a lens model for MG 0751+2716. The lens model, Model 1 (shown in Table \ref{model1tab}), from \citealt{2018MNRAS.478.4816S} was used to fit our data using the {\sc visilens} package. The images from left to right correspond to the dirty image of the lensed source, the dirty image of the lensed source obtained by fitting the lens model, the corresponding residual map, a high-resolution model, and the source plane model.}
    \label{modelcont0751}

\end{figure*}

To ensure that the MCMC walkers explore the parameter space well, we ran the  optimisation several times, starting  with a different set of initial parameter values every time. The different MCMC runs gave very similar sets of parameters and, consequently, very similar looking lensed images from the fitted model. One such example is shown in Fig. \ref{modelcont0751}. The range of the final parameter values found from the runs is shown in Table \ref{model1tab}.

The initial values for the source major axis in the different MCMCs were 500 mas, 100 mas, 80 mas, and 50 mas. We found that in all these cases, the optimisation converged to
a source full width at half maximum (FWHM) of around 30 to 60 mas. At the redshift of the source, this corresponds to 225 to 450~pc. Most of the chains converged towards a FWHM of 60 mas (450~pc). Figure 12 of \cite{2020MNRAS.495.2387S} demonstrates that most of the emission from the 27.4 GHz radio source originates from an approximately 500 pc region. We also see that the total size of the radio source in this study is approximately equal to 0\farcs3, which corresponds to around 2~kpc at this redshift, a figure that is consistent with the size of the radio source studied in \cite{2020MNRAS.495.2387S}. This also shows that the quasar radio structure corresponds to an extended, elliptical source (under the assumption of a single elliptical Gaussian source model).

Several studies of MG 0751+2716 at   radio and   optical wavelengths have been conducted in the past \citep{2005ApJ...618..586C,2006ApJ...649..635R,2018MNRAS.478.4816S,2020MNRAS.495.2387S,2021MNRAS.501..515P}. Our results show consistency with \citealt{2020MNRAS.495.2387S}; this study finds, at high  radio frequencies (27~GHz), radio AGN jets embedded in the extended molecular gas. This consistency implies an absence of an additional steep-spectrum component emitted in the radio; the low-frequency radio source is co-spatial with the core-jet structure that forms the radio structure at higher frequencies, and no significant lobe emission or further components associated with star formation are visible within the magnified region of the lens.  


\section{\class}

\subsection{The lens system B1600+434}

\begin{figure}
\centering
\includegraphics[scale=0.6]{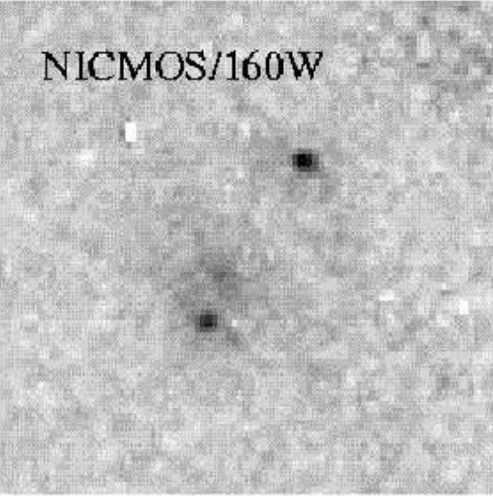}
\caption{HST infrared (1.6~\textmu m) image of CLASS B1600+434. The image is 3\farcs69 on a side and the scale is 43 mas/pixel. Image reproduced from \citet{2000MNRAS.311..389J}.}
\label{nicmos1600}
\end{figure}

\class\ (16$^{\rm h}$01$^{\rm m}$40.45$^{\rm s}$ $43^{\circ}16^{\prime}47\farcs78$) is a gravitational lens system that was discovered during the Cosmic Lens All-Sky Survey \citep{1995MNRAS.275.25-29,2003MNRAS.341...13B,2003MNRAS.341....1M}. The source is lensed into two images at both radio and optical frequencies, with a separation of 1\farcs4 and a flux ratio of $1.30 \pm 0.04$ at 8.4~GHz \citep{1995MNRAS.275.25-29}. The main lens is an edge-on spiral galaxy at $z=0.41$, and the source is a quasar at $z=1.59$ \citep{1997A&A.317.39-42,1998AJ....115..377F,1998MNRAS.295..534K}. The fainter  south-eastern image (image B) is close to the line of sight through the lensing galaxy, and appears reddened in the optical by passage through  the lens, with a differential reddening between image B and A of about 1 magnitude in the optical V band \citep{1997A&A.317.39-42}. This is confirmed by infrared Hubble Space Telescope (HST) images at 1.6~\textmu m \citep{2000MNRAS.311..389J}, which reveal that the infrared image A/B flux ratio is indistinguishable from that seen in the radio (Fig.~\ref{nicmos1600}). The main aim of observing \class\ is to use the two lines of sight from the same background object, corresponding to images A and B, to assess whether there is any evidence at these low radio frequencies for environmental effects, namely scatter broadening or free-free absorption in the lensing galaxy. These effects would result in an increase in size, or reduction in flux, of the B image, which passes close to the lensing galaxy. Recently  upper limits have been derived \citep{2021MNRAS.505.2610B}, using VLBI at GHz frequencies, for environmental effects. Although our resolution is approximately a factor of 300 worse than this study, the 10-30 times lower frequency in this work combined with the $\sim\nu^{-2}$ dependence of absorption and scattering effects mean that we can derive limits of similar stringency.

\subsection{Observations and data reduction}

The HBA was used to observe a bright calibrator source 3C 196 at a frequency range 120 to 183 MHz on 2015 September 30 for 10 minutes. Following this, observations of \class\ were taken on the same date for 8 hours at the same frequency range as the calibrator source. All core stations, remote stations within the Netherlands, and international stations that were then available (i.e. all currently available international stations except the Polish and Irish stations) were used for the observations. Observations of this object were carried out as part of the LOFAR Surveys Key Science Project (LSKSP; \citealt{2017A&A...598A.104S}) with a pointing centre of $16^{\rm h}03^{\rm m}32.88^{s}  ~42^{\circ}33^{\prime}21^{\prime\prime}$. This pointing centre is about $0.9^{\circ}$ from the target. At this radius, we expect an amplitude reduction of approximately a factor of 2-4 due to a combination of integration time smearing and bandwidth smearing, with the integration time and channel width used in the observation.

To calibrate the Dutch stations, {\sc prefactor} was carried out using the Offringa high-resolution sky model\footnote{Available from \url{https://github.com/lofar-astron/prefactor/blob/master/skymodels/3C196-offringa.skymodel}} for 3C~196, and using CS001 for the reference station. 
With the calibrator source, all core and remote stations produced good solutions for TEC and bandpass except for CS024. For the target source, these again provided good results except for stations RS205, RS208, and RS306. These stations and corresponding baselines were flagged during further analysis. In addition, data above 170 MHz were discarded due to the phase between the XX and YY polarisation calibrations being significantly large (much greater than 1 radian for a major part of the observation); these data are likely to be severely affected by RFI.

Following {\sc prefactor}, the LOFAR-VLBI Pipeline \citep{paper1} was used for the sub-arcsecond data calibration. L256173 (15$^{\rm h}$59$^{\rm m}$30.92$^{\rm s}$ 43$^{\circ}$49$^{\prime}$15.80$^{\prime\prime}$, 4C+43.36) was selected as a calibrator from the LBCS survey. This source is approximately 1.5$^{\circ}$ from the field centre, and is likely to be subject to greater amplitude reductions than the target due to smearing effects. 

For calibration of this source, the data was preprocessed using the Default Preprocessing Pipeline (DPPP). To obtain good solutions for TEC, direction-dependent gains were calibrated using the DDECal step in DPPP \citep{van_diepen_dppp_2018}. The solutions were smoothly varying, and values were below 2 TECU for all international stations except UK608 and DE609; these stations did not produce good solutions. The output of this process was phase shifted and averaged by factors of 8 in frequency and time, producing small datasets of both L256173 and \class\ at a frequency resolution of two channels per unflagged sub-band or a total of 360 channels.

\begin{figure}
    \includegraphics[width=8cm]{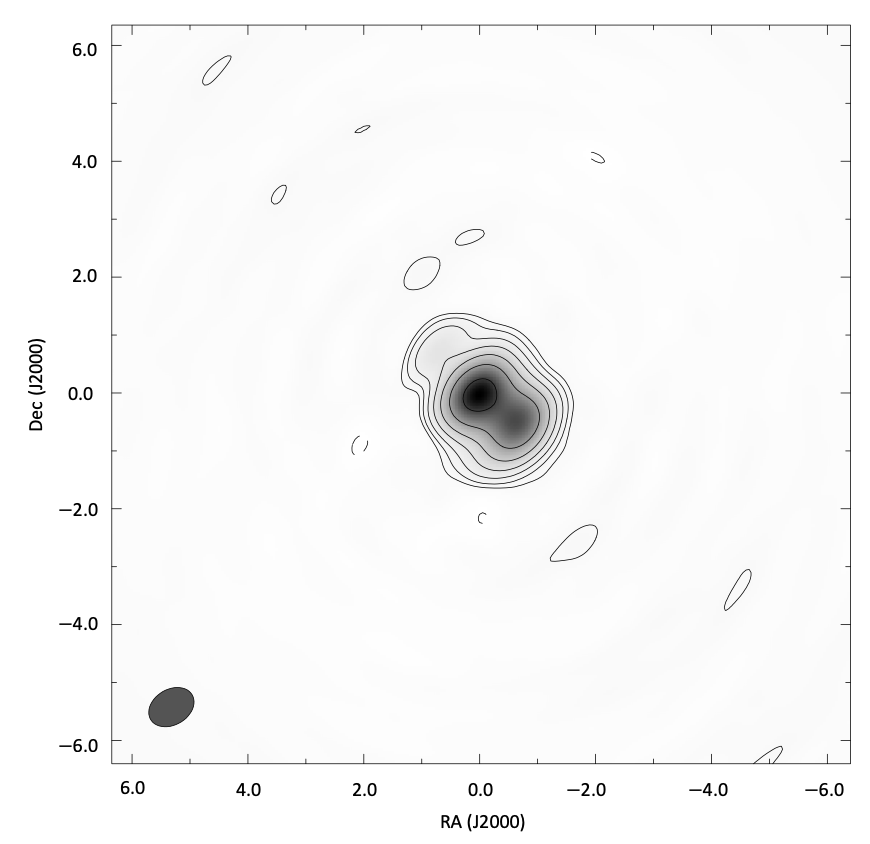}
    \caption{International LOFAR Telescope image of L256173 at 140 MHz, with a restoring beam of 0\farcs83$\times$0\farcs61 in PA $-59.1^{\circ}$. The grey-scale runs from 0.0 to 0.7~mJy/beam and the contours are at 0.00786$\times$($-$1,1,2,4,8,26,32,64) mJy\,beam$^{-1}$. }
    \label{map1559}
\end{figure}

\begin{figure}
    \includegraphics[width=8cm]{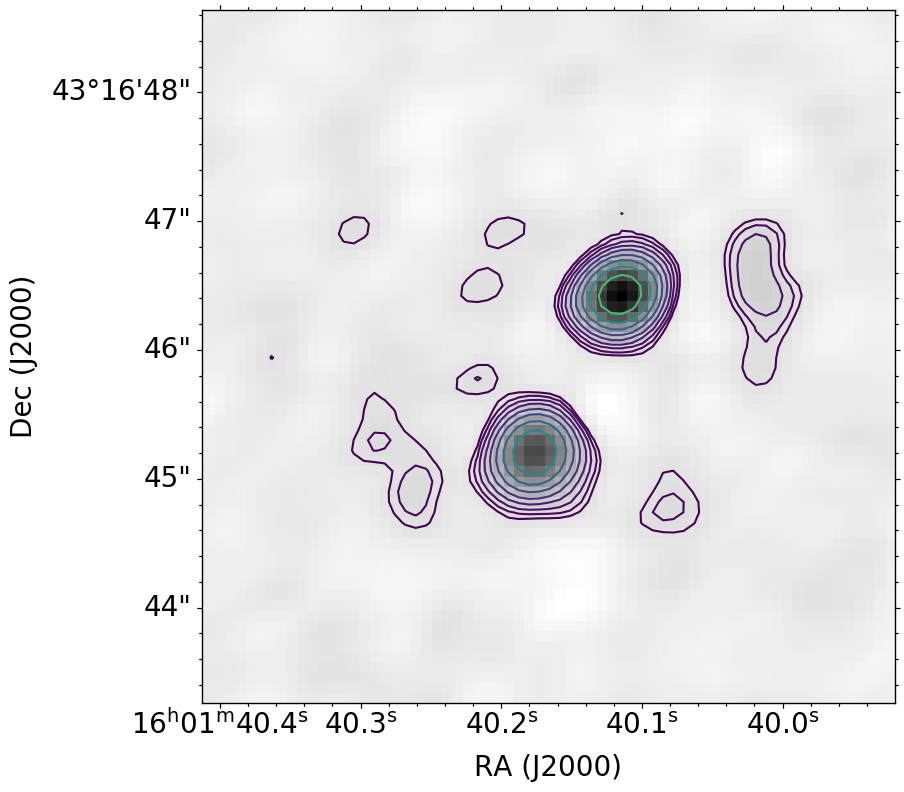}
    \caption{International LOFAR Telescope image of \class  ~at 140 MHz, with a restoring beam of 0\farcs46$\times$0\farcs30 in PA $-61.4^{\circ}$.  The north-western image is A, while B is the south-eastern one. The lowest contour is at 1.4~mJy/beam, and contours increase in steps of $\sqrt{2}$.}
    \label{map1600}
\end{figure}

The calibrator L256173 was successfully mapped with the pipeline (Fig. \ref{map1559}) and was found to be extended by $\sim$1\farcs5. For \class, it proved difficult to obtain good delay and phase solutions from the standard pipeline. Its output had jumps in phase at regular intervals in frequency. Accordingly, the data, with the initial corrections from DDECal, were read into the Astronomical Image Processing System (AIPS)\footnote{Distributed by the US National Radio Astronomy Observatory: http://aips.nrao.edu} and separated into 19 IFs with 40 channels each, which corresponded to the interval between each jump. A successful calibration for phase and delay was obtained with {\sc fring} using a one-minute solution interval for both delay and phases, a signal-to-noise threshold of 2 for solutions, a delay window of 400 ns, and a previously made map from a previous iteration of the mapping procedure. The procedure yielded good solutions, which were edited and smoothed before application to the target source. Data were averaged to 1 minute in time and 196 kHz channels and imaged using robust 0 weighting using only $>$80 km baselines. They were then phase self-calibrated with a ten-minute solution interval and using one phase solution for the whole band (because of the faintness of B1600+434 there was not a sufficient signal-to-noise ratio on less than that), together with a model consisting of the clean components from an initial image. Images with the phase self-calibrated dataset were produced (Fig.~\ref{map1600}).  

The initial flux density scale was produced by the {\sc prefactor} pipeline. It is more difficult to establish for this object because of the large distances of the source and phase calibrator from the LoTSS field centre, which is likely to result in significant flux loss due to time- and bandwidth-smearing. The phase calibrator L256173 has a flux density of 3.67~Jy in a low-resolution survey at 151~MHz \citep{1988MNRAS.234..919H} compared to 2.3~Jy at the same frequency here. 

The flux density scale was therefore established using data from the 6$^{\prime\prime}$ resolution images from the Dutch-only stations,   part of the main LoTSS survey \citep{2017A&A...598A.104S}, in which a total flux density of 57~mJy is measured; images were scaled to match this measurement. In the scaled image, the A and B flux densities from this LOFAR observation were measured using {\sc jmfit} to be 33.2$\pm$0.5 and 27.9$\pm$0.5~mJy with errors from the fit only and excluding systematic error on the flux calibration. The values are consistent with the Westerbork Northern Sky Survey (WENSS) overall flux density of 40-50~mJy at 325~MHz \citep{rengelink97a}, although this object is known to be variable \citep{2003ApJ...595..712K}. The final map has a noise level of about 220\,\textmu Jy/beam far from the sources, and is shown in Fig. \ref{map1600}. The noise increases close to the source, due to residual uncalibrated phase and/or amplitude errors; the signal-to-noise level does not allow for the removal of short-timescale phase variations or residual amplitude errors.

\subsection{Propagation effects}

The flux ratio of images A and B in \class\ is 1.19$\pm$0.04 in these 150 MHz LOFAR observations, which is statistically indistinguishable from the ratio at 8.5 GHz (\citealt{2021MNRAS.505.2610B}; the flux ratio found in this study varied from 1.19 to 1.26, depending on the observing season), and  indistinguishable from the predicted ratio of 1.25$\pm$0.03 for mass models of the system \citep{1998MNRAS.295..534K}. This is despite the fact that the light path of image B passes close to the centre of the edge-on spiral galaxy that acts as the lens.

Two propagation effects are capable of affecting radio waves passing through ionised gas in the lensing galaxy: free-free absorption (e.g. \citealt{2007A&A...465..405M, 2003ApJ...587...80W}) and scattering. In case of free-free absorption the optical depth is given by $\tau = 0.08235~ T_e^{-1.35} \nu^{-2.1} E$, where $T_e$ is the electron temperature in Kelvin, $\nu$ is the frequency in GHz, and $E$ is the emission measure in cm$^{-6}$ pc of the ionised medium \citep{1967ApJ...147..471M}. The fact that the image flux ratios in this system are the same as those at higher frequencies (\citealt{2021MNRAS.505.2610B}) implies an optical depth $\tau\ll 1$ at the rest frequency of the emission from the lens galaxy. This in turn implies $E\ll 42000$~cm$^{-6}$ pc for typical $T_e\sim 5000$~K, which is unsurprising given a passage through a typical galactic column of $\sim$10~kpc with $n_e\sim 0.03$~cm$^{-3}$, but much less than those in {\sc Hii} regions in the CLASS~B0218+357 lens galaxy \citep{2007A&A...465..405M}.  

The theory of scattering is discussed by a number of authors \citep{1977ARA&A..15..479R,2001Ap&SS.278..149W,2001ApJ...549..997C,2003MNRAS.338..599B,2007A&A...465..405M,2016MNRAS.459.2394Q}; its effect is to broaden point-source background structures with a characteristic width, $\theta$, of

\begin{equation}
\left(\frac{\theta}{1 {\rm mas}}\right) =  64\ {\rm SM}^{3/5}\left(\frac{\nu^{\prime}}{1 {\rm GHz}}\right)^{-11/5},
\end{equation}

\noindent where $\nu^{\prime}$ is the frequency at the redshift of the scatterer, and SM is the scattering measure in units of kpc\,m$^{-20/3}$ \citep{2001Ap&SS.278..149W}. Estimates of scattering measure have typically been derived for lenses using VLBI observations at GHz frequencies. For example, broadening of a few milliarcseconds in the image of CLASS~B0218+357, which passes through a molecular cloud in the lensing galaxy, implies SM$\sim$100~kpc\,m$^{-20/3}$ \citep{2003MNRAS.338..599B,2007A&A...465..405M}, and similar results are derived in CLASS~B0128+437 by \cite{2004MNRAS.350..949B} and other lens systems \citep{1996ApJ...470L..23J,1999MNRAS.305...15M,2003A&A...405..445W}.

Using {\sc jmfit}, we fitted two elliptical Gaussians at the positions of the images and found the deconvolved widths of B and A to be 386.83~mas and 287.8~mas respectively. Consequently, we find an upper limit of 258.5 mas on the value of  $\theta$. This corresponds to a differential SM$\lesssim$0.035 kpc\,m$^{-20/3}$; considering a lens galaxy redshift of 0.41 implies $\nu^{\prime}=212$~MHz. Here, we assume that image A is not affected by scattering. The line of sight of image B passes 250$\pm$50 mas (1.4$\pm$0.3~kpc) from the centre of the edge-on spiral lens \citep{2000MNRAS.311..389J}, about 1~kpc above its plane. A study of \class ~has also been carried out by \cite{2021MNRAS.505.2610B} using higher resolution observations, but at a higher frequency; these also yield an upper limit on the presence of scatter broadening in the system. Assuming an upper limit of 1.67 mas on $\theta$ at an observing frequency of 1.4 GHz (based on the sizes of A and B found by \citealt{2021MNRAS.505.2610B} at 1.4 GHz), we find an upper limit of SM$\lesssim$0.028~kpc\,m$^{-20/3}$, a value approximately equal to the one we found in this study.

The most direct comparison available is with our own Galaxy. Extensive modelling of Galactic free electrons has been done using pulsar studies \citep{2001ApJ...549..997C}, and VLBI observations of extragalactic point sources seen through the bar of our Galaxy also imply extensions of about 1~mas \citep{2015MNRAS.452.4274P} at observing frequencies of 2 and 8 GHz, implying scattering measures of about 1~kpc\,m$^{-20/3}$ along this line of sight. Scatter-broadening is not confined to milliarcsecond scales, however: an extreme scattering event in NGC~6334 \citep{1998ApJ...493..666T} results from the passage of a point background radio source through an area of molecular clouds and H{\sc ii} regions in our Galaxy. At 20~cm a consequent $3^{\prime\prime}$ broadening is observed, corresponding to SM$>$1000~kpc\,m$^{-20/3}$ in an exceptional line of sight.

\section{Unsuccessful sources}

\subsection{MG~1549+3047: Observations and data reduction}

The gravitational lens system MG~1549+3047 ($15^{\rm h}49^{\rm m}12.6^{\rm s} +30^{\circ}47^{\prime}15^{\prime\prime}$) was also discovered as part of the MG survey of radio sources, and consists of a lensed radio lobe in a radio galaxy \citep{1993AJ....105..847L} at redshift 1.17 \citep{2003MNRAS.343L..29T} imaged by a foreground galaxy at $z=0.11$ \citep{1996AJ....111.1812L}.

\begin{figure}
    \includegraphics[width=8cm]{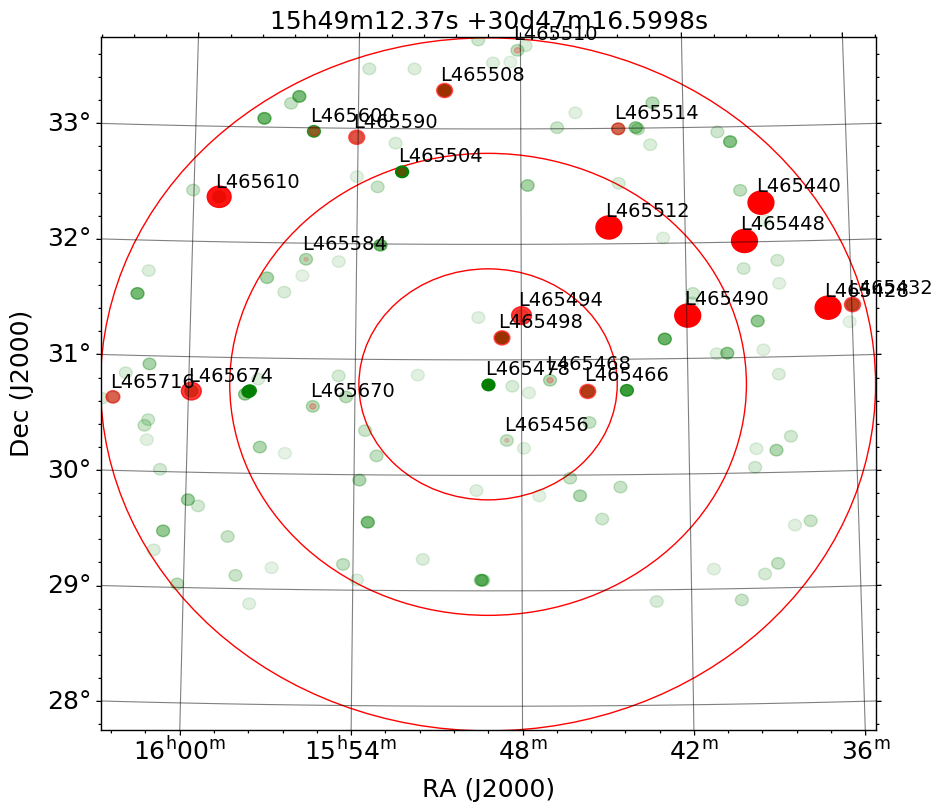}
    \caption{LBCS sources in the neighbourhood of MG1549+3047. Sources from the Westerbork Northern Sky Survey (declination limit of about 29$^{\circ}$ are in green and LBCS sources are in red.}
    \label{lbcs1549}
\end{figure}

Figure~\ref{lbcs1549} shows the LBCS map of the field. The source itself was observed with LBCS and was found to have significant correlated flux in the three minutes of integration used for LBCS only on the shortest international baselines (from the array centre to DE609 Norderstedt, with some correlated flux on the other German baselines). The two nearest good calibrators lie within one degree. The farther one, L465494, gave a good correlated signal on all international baselines (although the PL610, PL611, PL612, and IE613 stations were not operational at the time of the observations), and the nearer one, L465498, gave a good signal on the shorter baselines (200-300~km).

MG~1549+3047 was observed on 2018 March 13 for a standard eight-hour track, with short ten-minute calibration scans of 3C196 and 3C295 on either side of the observation. The Dutch stations were calibrated as described in the previous section using the {\sc prefactor} procedures using the 3C295 calibration scan, together with the high-resolution model of 3C295 (credit: F. Sweijen\footnote{https://github.com/lofar-astron/prefactor}). The RS306 station was again excluded from the dataset. Bandpass and TEC solutions were inspected and found to be satisfactory, with coherent phase solutions derived on the Dutch stations from initial calibration of the target field against the TGSS data.

The field source L465494 was used to solve for the delays. In this case, and unlike the case of \mg, satisfactory delay solutions could not be obtained unless short baselines (those which did not involve an international station) were excluded; in this case, smoothly varying TEC solutions were derived for all international stations. Smaller datasets, averaged in time and frequency were again derived for the target, for the field calibrator sources within 1 degree (L465494, L465498 and L465466, see Fig.~\ref{lbcs1549}), and for a brighter but more distant calibrator source (L465490). However, tests showed that in this case, it was not possible to transfer phase solutions satisfactorily from the calibrator to the target or to other calibrator sources. Some signal can be recovered by self-calibration on the target itself, but the recovered map is highly dependent on the assumed starting model.

\subsection{Cycle 0 observations}

Two further sources were observed in 2013 as part of Cycle 0, after the first LOFAR call for proposals: Q0957+561 \citep{1979Natur.279..381W} and MG~1131+0546 \citep{1988Natur.333..537H}. These observations suffered from significant problems, the main one being the quality of station calibration for the international stations which was not of the same standard as the current calibration. This resulted in very low signal on the calibrator for some international stations, and a consequent failure to fit delay solutions. It is likely that good international-baseline maps cannot be produced from data taken this early. Q0957+561 is now available as part of the LSKSP field, and this object is therefore under investigation with the new data.

\section{Conclusions}

Two radio-loud gravitational lenses, \mg\ and \class , have been observed with the LOFAR HBA at a frequency centred around 150~MHz. We obtained the first high-resolution images of lens systems at such a low frequency with the ILT. The low-frequency structure of the recovered source in \mg\ is very similar to that at higher frequencies, and no evidence is found for extra steep-spectrum radio-emitting components. In \class\ we find a flux ratio between the double images of the background quasar that is consistent with the value at higher frequencies. The widths of the two components give a consequent limit on the differential scattering measure, and the flux ratios give a limit on the electron density and/or clumpiness of two lines of sight through the bulge of the edge-on spiral lens galaxy. Attempts at calibrations of three further observations, two of them being Cycle 0 observations, were unsuccessful. Further investigations are under way to improve the reliability of the imaging pipeline, and to understand the possible effects of the varying ionosphere on the calibratibility of some observations.

\begin{acknowledgements}
     We thank the anonymous referee for their comments on the paper. This paper is based on data obtained with the International LOFAR Telescope (ILT) under project code LC9-012. Part of this work was supported by LOFAR, the Low Frequency Array designed and constructed by ASTRON, that has facilities in several countries, that are owned by various parties (each with their own funding sources), and that are collectively operated by the ILT foundation under a joint scientific policy. Data analysis for this project used the LOFAR-UK computing cluster based at the University of Hertfordshire. JPM acknowledges support from the Netherlands Organization for Scientific Research (NWO) (Project No. 629.001.023) and the Chinese Academy of Sciences (CAS) (Project No. 114A11KYSB20170054). LKM is grateful for support from the UKRI Future Leaders Fellowship (grant MR/T042842/1).
\end{acknowledgements}
\bibliographystyle{aa}

\bibliography{lofarlens.bib}

\end{document}